# Sunspot latitudes during the Maunder Minimum: a machine-readable catalogue from previous studies

José M. Vaquero[1*], José M. Nogales[2] and Florentino Sánchez-Bajo[3]

[1]Departamento de Física, Centro Universitario de Mérida, Universidad de Extremadura, Avda Santa Teresa de Jornet 38, 06800 Mérida, Spain, jvaquero@unex.es

[2]Departamento de Expresión Gráfica, Centro Universitario de Mérida, Universidad de Extremadura, Avda Santa Teresa de Jornet 38, 06800 Mérida, Spain, jmnogale@unex.es

[3]Departamento de Física Aplicada, Escuela de Ingenierías Industriales, Universidad de Extremadura, Avda de Elvas s/n, 06006 Badajoz, Spain, fsanbajo@unex.es

## Abstract

The Maunder Minimum (1645-1715 approximately) was a period of very low solar activity and a strong hemispheric asymmetry, with most of sunspots in the southern hemisphere. In this paper, two data sets of sunspot latitudes during the Maunder minimum have been recovered for the international scientific community. The first data set is constituted by latitudes of sunspots appearing in the catalogue published by Gustav Spörer nearly 130 years ago. The second data set is based on the sunspot latitudes displayed in the butterfly diagram for the Maunder Minimum which was published by Ribes and Nesme-Ribes almost 20 years ago. We have calculated the asymmetry index using these data sets confirming a strong hemispherical asymmetry in this period. A machine-readable version of this catalogue with both data sets is available in the Historical Archive of Sunspot Observations (http://haso.unex.es) and in the appendix of this article.

## 1. Introduction

The Maunder Minimum (hereafter, MM) is a special episode of the Sun history (Eddy, 1976) when sunspots almost completely vanished, while the solar wind kept blowing, although at a reduced pace (Cliver et al., 1998; Usoskin et al., 2001). Sunspots rarely appeared but an analysis of $^{10}$Be data (Beer et al., 1998) implies that the 11-year cycle was weak but fairly regular during the MM. Reviews about the MM and the historical aspect of solar activity can be found in Soon and Yaskel (2003), Vaquero (2007), Vaquero and Vázquez (2009) and Usoskin (2013). Although the classical scenario implies that the transition from the normal high activity to the deep minimum did not have any apparent precursor, newly recovered historical observations suggest that the onset of the MM could not have been very sudden (Vaquero et al., 2011). Other important feature of sunspot activity during the MM was a strong north-south (hereafter, N-S) asymmetry. Sunspots were mainly observed in the southern solar hemisphere (Spörer, 1889; Ribes and Nesme-Ribes, 1993).

---

[*]Corresponding author: José M. Vaquero, e-mail: jvaquero@unex.es, tel.: 0034 924 289 300, fax: 0034 924 301 212

Two important studies have clearly shown this last feature of the MM. On the one hand, Spörer (1889) indicated, in a pioneering paper published in the 19th century, this strong asymmetry, listing the known sunspot positions, based mainly on published observations and drawings. Furthermore, Ribes and Nesme-Ribes (1993) reported the characteristics of solar activity during the MM from observations preserved in the archives of the Paris Observatory. Moreover, they published a butterfly diagram for the 1670-1720 period based on these observations. These two studies are the main source of sunspot latitudes during the MM except the original observations (Clette et al., 2014; Casas et al., 2006).

Although several butterfly diagrams have been made in recent years with data from 18th (Arlt, 2009; Cristo et al., 2011) and 19th (Arlt et al., 2013; Casas and Vaquero, 2014) centuries, no data are available (in a machine-readable version) to build a butterfly diagram for the 17th century (especially for MM). In regard to this, the aim of this paper is to provide the international scientific community a machine-readable version of sunspot latitudes from the works of Spörer (1889) and Ribes and Nesme-Ribes (1993).

## 2. Data and Method

As we have pointed out in the preceding section, we have recovered data from the works of Spörer (1889) and Ribes and Nesme-Ribes (1993). The origin of the data is quite different. Spörer (1889) provided only a list of sunspot latitudes. However, Ribes and Nesme-Ribes (1993) consulted the original data in the archive of the Paris Observatory and published a butterfly diagram (upper panel of Figure 6 in Ribes and Nesme-Ribes, 1993). Therefore, we have built a database extracting the information of both sources using different methods.

It was very easy to copy the data provided by Spörer (1889). In total, 64 sunspot latitudes associated with dates were listed by this author. Graphical data provided by Ribes and Nesme-Ribes (1993) were converted to a machine-readable file using the web interface called *WebPlotDigitizer* (http://arohatgi.info/WebPlotDigitizer/) that have been elaborated by Ankit Rohatgi. This code has been developed to facilitate the process of data extraction in an easy and accurate way from a variety of plot types. This program was built using HTML5. Therefore, it requires no installation on to the hard drive of the user, running within a browser. In total, using *WebPlotDigitizer*, 213 sunspot latitudes associated with dates were obtained from the data of the Figure 6 of Ribes and Nesme-Ribes (1993).

Note that all the observers included in the present two data sets are included in the data set collected by Hoyt and Schatten (1998). Spörer (1889) indicated that he used the bibliography on sunspots that appeared in the journal "Mittheilungen über die Sonnenflecken", by R. Wolf. This journal was extensively used by Hoyt and Schatten (1998), specially for the historical period and, obviously, including the MM period. Thus, well-known sunspot observations during the MM by Lalande, Cassini, Flamsteed, Halley, De la Hire, Wurzelbaur, Maraldi, Manfredi, Derham, Wiedenburg and Kirch were compiled by Spörer (1889). Moreover, Ribes and Nesme-Ribes (1993) used the historical sunspot observations preserved in the archive of the Paris Observatory. Hoyt and Schatten (1998) acknowledged in their bibliography the data from Elizabeth Nesme-Ribes of the French astronomers. For example, in the references about Phillippe de La Hire, one of the most important sunspot observers in the MM period, they stated: "Primary source is re-examination of original notebooks by Elizabeth Nesme-Ribes at the Paris Observatory". Therefore, it is clear that all the present original observers are included in the data set collected by Hoyt and Schatten (1998).

It is awkward to associate sunspots appearing in both sets of data due to errors in dating and latitude of each spot. In the Figure 6 of Ribes and Nesme-Ribes (1993), each sunspot is marked by a small square. This symbol indicates the date on the horizontal axis and the latitude of the spot on the vertical axis. If one compares the dimensions of this small square with the dimensions of the axes, each small square is about 1.1 years wide and 1.3 degrees high. The maximum error in determining the center of the square can be approximately 2/3 of these dimensions. Therefore, it is only possible to define the coordinates of the spot with accuracies of ± 0.7 years and ± 0.9 degrees. Thus, it is not simple to establish whether a given sunspot in a data set is or not repeated in the other one. To do this, the original observations should be used.

## 3. Results

We have merged both datasets into a single file containing three columns. The first column indicates the date associated to the sunspot latitude, that is listed in the second column. We have used the criterion of assigning the date of each sunspot to the midpoint between the extreme observation dates indicated by Spörer (1889). The last column shows the source. Figure 1 shows the sunspot latitudes from Spörer (red circles) and Ribes and Nesme-Ribes (blue squares).

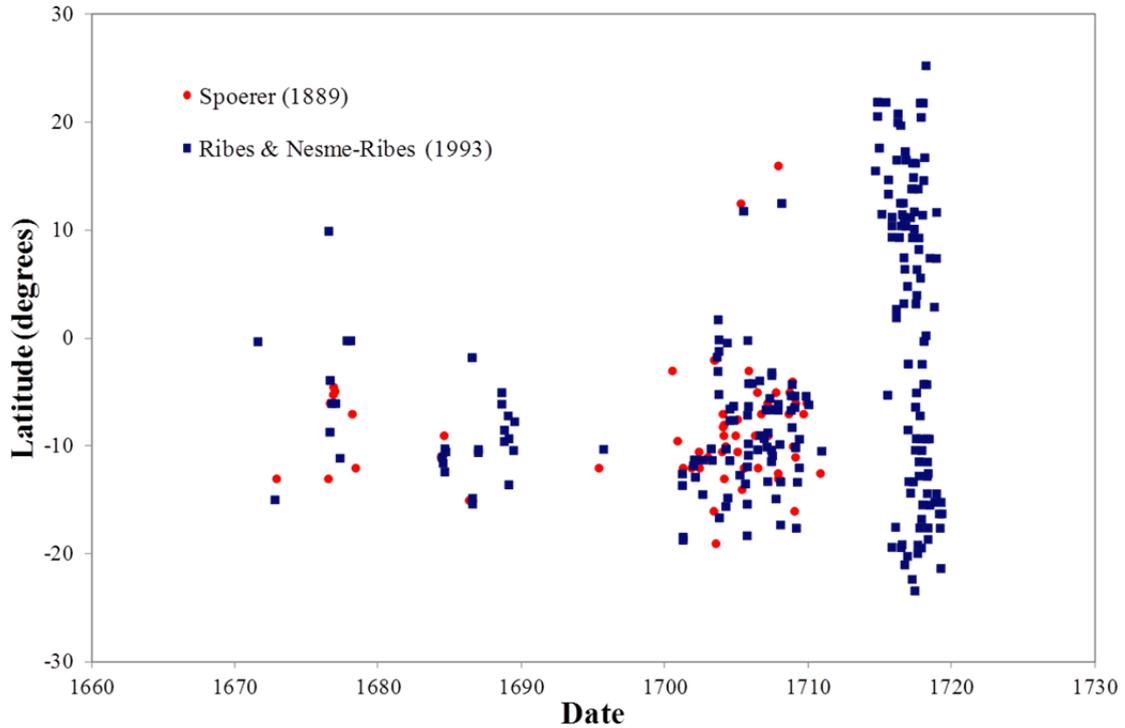

**Figure 1**. Butterfly diagram using the data from Spörer (red circles) and Ribes and Nesme-Ribes (blue squares).

Since the purpose of this work is to make the data available and archived, a copy of this single file is freely available at http://haso.unex.es and printed version appears in the appendix of this article.

## 4. Asymmetry through the MM

Both data sets show clearly the strong N-S asymmetry during the MM with sunspots mainly observed in the southern solar hemisphere. In the first solar cycle after the MM in the 1710s this asymmetry vanished. For studying the N-S asymmetry in each decade, we have calculated the asymmetry index $A$, defined as (Verma, 1993)

$$A = \frac{S_N - S_S}{S_N + S_S}$$

$S_N$ and $S_S$ being, respectively, the number of spots observed in the northern and southern hemispheres. Table 1 displays the values of $S_N$ and $S_S$, $S_N + S_S$ and the index $A$ in each decade for the Spörer and Ribes and Nesme-Ribes data sets, besides of the whole data set. It is clear that, except for the Ribes and Nesme-Ribes data corresponding to the 1711-1720 decade, the data exhibit a strong N-S asymmetry. Moreover, the asymmetry of the data was analyzed from the point of view of the probability of having such an asymmetric distribution by random fluctuation. This can be done using the cumulative binomial probability (Press et al. 1989; Chowdhury and Dwivedi, 2011)

$$P = \sum_{j=k}^{n} \binom{n}{j} p^j (1-p)^{n-j}$$

that, in this context, provides the probability of having a number of sunspots equal or greater that $k$ in the southern hemisphere ($n$ and $p$ being, respectively, the total number of sunspots and the random probability $-p = 0.5$ in our case- of having a sunspot in the southern hemisphere). Above equation can be expressed in terms of the incomplete beta function (Abramowitz and Stegun, 1972)

$$P = I_p(k, n-k+1)$$

**Table 1**. Information on sunspots observed in the northern and southern solar hemispheres in the different decades in the two data sets: hemispheric and total number of sunspots, the asymmetry index and the probability of having the observed asymmetric distribution by random fluctuations.

| Data set | $S_N$ | $S_S$ | $S_N + S_S$ | $A$ | $P(\%)$ |
|---|---|---|---|---|---|
| Spörer 1671-1680 | 0 | 8 | 8 | -1 | 0.4 |
| Spörer 1681-1690 | 0 | 4 | 4 | -1 | 6.3 |
| Spörer 1691-1700 | 0 | 3 | 3 | -1 | 12.5 |
| Spörer 1701-1710 | 2 | 47 | 49 | -0.918 | $2.2\ 10^{-10}$ |
| Ribes and Nesme-Ribes 1671-1680 | 1 | 9 | 10 | -0.8 | 1.1 |
| Ribes and Nesme-Ribes 1681-1690 | 0 | 19 | 19 | -1 | $1.9\ 10^{-4}$ |
| Ribes and Nesme-Ribes 1691-1700 | 0 | 1 | 1 | -1 | 50 |
| Ribes and Nesme Ribes 1701-1710 | 3 | 73 | 76 | -0.921 | $9.7\ 10^{-17}$ |
| Ribes and Nesme-Ribes 1711-1720 | 58 | 49 | 107 | 0.084 | 83.3 |
| Complete data set 1671-1680 | 1 | 17 | 18 | -0.889 | $7.2\ 10^{-3}$ |
| Complete data set 1681-1690 | 0 | 23 | 23 | -1 | $1.2\ 10^{-5}$ |
| Complete data set 1691-1700 | 0 | 4 | 4 | -1 | 6.3 |
| Complete data set 1701-1710 | 5 | 120 | 125 | -0.92 | $5.7\ 10^{-28}$ |

In regard to this, table 1 shows also the corresponding values of $P$ calculated for the same data sets considered in the case of the asymmetry index.

Table 1 shows that, except for the Ribes and Nesme-Ribes data of the 1711-1720 decade, a noticeable N-S asymmetry is present, with values of the asymmetry index $A$ equal or very close to -1. At a significance level of 5%, the data of Spörer in the decades with a very low number of recorded sunspots (1681-1690 and 1691-1700) are no significant from the point of view of the asymmetry, although the 1681-1690 data can be considered marginally significant. In the 1671-1680 and 1701-1710 decades, at the same significance level, the N-S asymmetry is significant in Spörer's data. The Ribes and Nesme-Ribes data, owing to the greater number of recorded sunspots, shows that, except for the 1691-1700 decade (with only a recorded sunspot), the asymmetry is significant. Note that, for the complete data set, the asymmetry is very significant in all cases ($P < 1\%$), with the exception of the 1691-1700 decade, where the asymmetry can be considered only marginally significant ($P = 6.3\%$).

## 4. Conclusion

We have recovered two data sets of sunspot latitudes during the MM for the international scientific community. The first data set is based in Spörer (1889) and the second data set is based on the sunspot latitudes displayed in the butterfly diagram for the MM which was published by Ribes and Nesme-Ribes (1993). We used the web tool *WebPlotDigitizer* to extract the data from the figure of the butterfly diagram published by Ribes and Nesme-Ribes (1993). Thus, a machine-readable version of these data is available in the Historical Archive of Sunspot Observations (http://haso.unex.es). These data could be used for studies that do not require high accuracy in latitude and date.

It is worthwhile to mention to note that it is not clear how was operating the solar dynamo during the MM. The existence of the 11-y Schwabe cycle is mainly based on proxies like the cosmogenic isotopes, while the sunspot observations rather give evidence for a dominant 22-year cycle (see, for example, Usoskin et al., 2013). The situation only will be resolved with close scrutiny of the ancient observations of the Sun that are preserved in the historical archives and libraries.

## Acknowledgements

Support from the Junta de Extremadura (Research Group Grants GR10131 and GR10045), Ministerio de Economía y Competitividad of the Spanish Government (AYA2011-25945) and the COST Action ES1005 TOSCA (http://www.tosca-cost.eu) is gratefully acknowledged.

## References

Abramowitz, M., and Stegun, I.A. "Handbook of Mathematical Functions with Formulas, Graphs, and Mathematical Tables". Dover Publications, Inc., New York, 1972.

Arlt, R. "The solar butterfly diagram in the 18th century". Solar Phys. 255:143-153, 2009.

Arlt, R., R. Leussu, N. Giese, K. Mursula, Usoskin, I.G. "Sunspot positions and sizes for 1825-1867 from the observations by Samuel Heinrich Schwabe". Mon. Not. R. Astron. Soc. 433:3165-3172, 2013.

Beer, J., Tobias, S. and Weiss, N. "An Active Sun Throughout the Maunder Minimum", Solar Phys. 181: 237–249, 1998.

Casas, R., Vaquero, J. M. and Vázquez, M. "Solar Rotation in the 17th century". Solar Phys. 234:379–392, 2006.

Casas, R., Vaquero, J.M. "The Sunspot Catalogues of Carrington, Peters, and de la Rue: Quality Control and Machine-readable Versions". Solar Phys. 289:79–90, 2014.

Chowdhury, P., and Dwivedi, B.N. "A study of the north-south asymmetry of sunspot area during solar cycle 23". First Asia-Pacific Solar Physics Meeting. ASI Conference Series, Vol. 2: 197–201, 2011.

Clette, F., Svalgaard, L., Vaquero, J.M., and Cliver, E. W. "Revisiting the Sunspot Number. A 400-year perspective on the solar cycle" Space Science Reviews 186, 35– 103.

Cliver, E.W., Boriakoff, V. and Bounar, K.H. "Geomagnetic activity and the solar wind during the Maunder Minimum", Geophys. Res. Lett. 25: 897–900, 1998.

Cristo, A., Vaquero, J.M., Sánchez-Bajo, F. "HSUNSPOTS: a Tool for the Analysis of Historical Sunspot Drawings". Journal of Atmospheric and Solar-Terrestrial Physics 73:187-190, 2011.

Eddy, J. A. "The Maunder Minimum". Science 192:1189–1202, 1976.

Hoyt, D.V., Schatten, K.H., "Group sunspot numbers: a new solar activity reconstruction". Solar Phys. 179, 189–219, 1998.

Press, W.H., Flannery, B.P., Teukolsky, S.A. and Vetterling, W.T. "Numerical Recipes. The Art of Scientific Computing". Cambridge University Press, Cambridge, UK, 1989.

Ribes, J.C. and Nesme-Ribes, E. "The solar sunspot cycle in the Maunder minimum AD1645 to AD1715". Astronomy and Astrophysics 276:549–563, 1993.

Soon, W.W.-H. and Yaskell, S.H. "The Maunder Minimum and the Variable Sun-Earth Connection", World Scientific, Singapore; River Edge, NJ. 2003.

Spörer, G. "Ueber die periodicität der Sonnenflecken seit dem Jahr 1618". Nova Acta der Ksl. Leop.-Carol. Deutschen Akademie der Naturforscher 53(2): 281-324, 1889.

Usoskin, I.G., Mursula, K. and Kovaltsov, G.A. "Heliospheric modulation of cosmic rays and solar activity during the Maunder minimum", J. Geophys. Res. 106: 16039–16046, 2001.

Usoskin, I.G. "A History of Solar Activity over Millennia", Living Rev. Solar Phys. 10: 1-94, 2013.

Vaquero, J.M. "Historical sunspot observations: A review". Advances in Space Research 40:929–941, 2007.

Vaquero, J.M., Vázquez, M. "The Sun Recorded Through History". Springer, Berlin, 2009.

Vaquero, J.M., Gallego, M.C., Usoskin, I.G., Kovaltsov, G.A. "Revisited sunspot data: A new scenario for the onset of the Maunder minimum". The Astrophysical Journal Letters 731:L24 (4pp), 2011.

Verma, V.K. "On the North-South asymmetry of solar activity cycles". The Astrophysical Journal 403: 797-800, 1993.

**Appendix. Sunspot latitudes during the Maunder Minimum**

In this appendix, sunspot latitudes during the Maunder Minimum recovered in this paper are listed. The first column indicates the date and the second column indicates the sunspot latitude. Finally, the third column shows the source of the data. We use "1" and "2" to indicate Spörer (1889) or Ribes and Nesme-Ribes (1993), respectively.

| Date | Latitude (°) | Source |
| --- | --- | --- |
| 1672.87945 | -13 | 1 |
| 1676.49178 | -13 | 1 |
| 1676.60822 | -6 | 1 |
| 1676.83288 | -5.2 | 1 |
| 1676.85753 | -4.5 | 1 |
| 1676.96164 | -4.9 | 1 |
| 1678.16301 | -7 | 1 |
| 1678.39863 | -12 | 1 |
| 1684.3589 | -11 | 1 |
| 1684.50548 | -10.8 | 1 |
| 1684.56986 | -9 | 1 |
| 1686.32055 | -15 | 1 |
| 1695.37123 | -12 | 1 |
| 1700.49726 | -3 | 1 |
| 1700.86027 | -9.5 | 1 |
| 1701.24384 | -12 | 1 |
| 1701.84658 | -12 | 1 |
| 1702.35206 | -10.5 | 1 |
| 1702.39041 | -12 | 1 |
| 1702.98767 | -11 | 1 |
| 1703.38904 | -16 | 1 |
| 1703.40822 | -2 | 1 |
| 1703.47945 | -2 | 1 |
| 1703.52877 | -19 | 1 |
| 1703.9726 | -10.5 | 1 |
| 1704.02055 | -7 | 1 |
| 1704.03425 | -8.2 | 1 |
| 1704.08356 | -9 | 1 |
| 1704.10411 | -8 | 1 |
| 1704.11096 | -13 | 1 |
| 1704.21644 | -10 | 1 |
| 1704.90685 | -9 | 1 |
| 1705.04384 | -10.5 | 1 |
| 1705.04384 | -7.5 | 1 |
| 1705.27534 | 12.5 | 1 |
| 1705.34932 | -14 | 1 |
| 1705.48767 | -12 | 1 |
| 1705.51918 | -12 | 1 |
| 1705.83014 | -3 | 1 |
| 1706.26986 | -9 | 1 |
| 1706.42466 | -5 | 1 |
| 1706.46575 | -12 | 1 |
| 1706.69726 | -7 | 1 |
| 1706.91918 | -9 | 1 |
| 1707.13699 | -6 | 1 |
| 1707.1863 | -9 | 1 |
| 1707.2274 | -6.5 | 1 |
| 1707.36575 | -6.5 | 1 |
| 1707.72466 | -5 | 1 |
| 1707.87123 | -12.5 | 1 |
| 1707.87671 | -13 | 1 |
| 1707.88493 | 16 | 1 |
| 1707.95616 | -13 | 1 |
| 1708.62055 | -7 | 1 |
| 1708.68767 | -5 | 1 |
| 1708.88082 | -4 | 1 |
| 1708.89863 | -6.5 | 1 |
| 1708.91644 | -10 | 1 |
| 1709.02192 | -16 | 1 |
| 1709.0863 | -11 | 1 |
| 1709.09315 | -6 | 1 |
| 1709.65206 | -7 | 1 |
| 1709.86712 | -6 | 1 |
| 1710.82466 | -12.5 | 1 |
| 1671.57392 | -0.3 | 2 |
| 1672.76594 | -14.9 | 2 |
| 1676.52035 | 9.9 | 2 |
| 1676.60161 | -8.7 | 2 |
| 1676.61401 | -3.9 | 2 |
| 1676.73796 | -6 | 2 |
| 1676.99688 | -6 | 2 |
| 1677.30746 | -11.1 | 2 |
| 1677.78881 | -0.2 | 2 |
| 1678.04773 | -0.2 | 2 |
| 1684.42858 | -11 | 2 |
| 1684.49194 | -11.6 | 2 |
| 1684.61933 | -12.4 | 2 |
| 1684.62484 | -10.2 | 2 |
| 1684.68888 | -10.5 | 2 |
| 1686.52408 | -1.8 | 2 |
| 1686.5537 | -15.3 | 2 |
| 1686.55507 | -14.8 | 2 |
| 1686.95448 | -10.6 | 2 |
| 1686.95517 | -10.3 | 2 |
| 1688.58447 | -6.1 | 2 |
| 1688.58722 | -5 | 2 |
| 1688.76971 | -9.5 | 2 |
| 1688.77246 | -8.5 | 2 |
| 1689.03483 | -7.2 | 2 |
| 1689.08303 | -13.5 | 2 |
| 1689.09405 | -9.3 | 2 |
| 1689.41495 | -10.4 | 2 |
| 1689.48657 | -7.7 | 2 |
| 1695.69457 | -10.3 | 2 |
| 1701.18847 | -13.6 | 2 |
| 1701.19122 | -12.6 | 2 |
| 1701.24012 | -18.7 | 2 |
| 1701.2408 | -18.4 | 2 |
| 1701.97006 | -11.8 | 2 |
| 1702.03617 | -11.2 | 2 |
| 1702.09677 | -12.8 | 2 |
| 1702.61049 | -14.5 | 2 |
| 1702.61875 | -11.3 | 2 |
| 1703.20409 | -10.2 | 2 |
| 1703.26607 | -11.3 | 2 |
| 1703.61451 | -1.7 | 2 |
| 1703.6758 | -3 | 2 |
| 1703.6882 | 1.7 | 2 |
| 1703.73502 | -5.2 | 2 |
| 1703.74535 | -1.2 | 2 |
| 1703.74811 | -0.1 | 2 |
| 1703.77014 | -16.6 | 2 |
| 1704.22602 | -15.6 | 2 |
| 1704.23979 | -10.2 | 2 |
| 1704.33 | -0.4 | 2 |
| 1704.35754 | -14.8 | 2 |
| 1704.49596 | -11.3 | 2 |
| 1704.5056 | -7.6 | 2 |
| 1704.50835 | -6.5 | 2 |
| 1704.76453 | -7.6 | 2 |
| 1704.76797 | -6.3 | 2 |
| 1705.20456 | -12.7 | 2 |
| 1705.46211 | 11.8 | 2 |
| 1705.59088 | -13.5 | 2 |
| 1705.70795 | -18.3 | 2 |
| 1705.71552 | -15.3 | 2 |
| 1705.72448 | -11.9 | 2 |
| 1705.73687 | -7.1 | 2 |
| 1705.75478 | -0.2 | 2 |
| 1705.79196 | -10.8 | 2 |
| 1705.79472 | -9.8 | 2 |
| 1705.80367 | -6.3 | 2 |
| 1705.80918 | -4.2 | 2 |
| 1706.0681 | -4.2 | 2 |
| 1706.44065 | -10.3 | 2 |
| 1706.58664 | -3.9 | 2 |
| 1706.63829 | -9 | 2 |
| 1706.89652 | -9.3 | 2 |
| 1707.03287 | -6.6 | 2 |
| 1707.08865 | -10.1 | 2 |
| 1707.14512 | -13.2 | 2 |
| 1707.15683 | -8.7 | 2 |
| 1707.2918 | -6.6 | 2 |
| 1707.29455 | -5.5 | 2 |
| 1707.40887 | -11.4 | 2 |
| 1707.41162 | -10.3 | 2 |
| 1707.42952 | -3.4 | 2 |
| 1707.43021 | -3.2 | 2 |
| 1707.47497 | -10.9 | 2 |
| 1707.72357 | -14.9 | 2 |
| 1707.87438 | -6.6 | 2 |
| 1707.87576 | -6.1 | 2 |
| 1707.99558 | -9.8 | 2 |
| 1708.04103 | -17.3 | 2 |
| 1708.05136 | -13.3 | 2 |
| 1708.11815 | 12.5 | 2 |
| 1708.78062 | -6.6 | 2 |
| 1708.78406 | -5.3 | 2 |
| 1708.84122 | -8.2 | 2 |
| 1708.85155 | -4.3 | 2 |
| 1709.04023 | -6.4 | 2 |
| 1709.04299 | -5.3 | 2 |
| 1709.09532 | -10.1 | 2 |
| 1709.14077 | -17.6 | 2 |
| 1709.21652 | -13.3 | 2 |
| 1709.34943 | -12 | 2 |
| 1709.35631 | -9.3 | 2 |
| 1709.81976 | -5.3 | 2 |
| 1710.01189 | -6.1 | 2 |
| 1710.90711 | -10.4 | 2 |
| 1714.66428 | 15.5 | 2 |
| 1714.80682 | 20.6 | 2 |
| 1714.81026 | 21.9 | 2 |
| 1714.92871 | 17.6 | 2 |
| 1715.10706 | 11.5 | 2 |
| 1715.39285 | 21.9 | 2 |
| 1715.5168 | -5.2 | 2 |
| 1715.565 | 13.4 | 2 |
| 1715.56845 | 14.7 | 2 |
| 1715.80396 | -19.3 | 2 |
| 1715.8136 | 9.4 | 2 |
| 1715.81635 | 10.4 | 2 |
| 1715.81842 | 11.2 | 2 |
| 1716.0677 | -17.5 | 2 |
| 1716.11797 | 1.9 | 2 |
| 1716.12004 | 2.7 | 2 |
| 1716.15585 | 16.6 | 2 |
| 1716.22953 | 20 | 2 |
| 1716.2316 | 20.8 | 2 |
| 1716.33145 | 9.4 | 2 |
| 1716.40444 | 12.6 | 2 |
| 1716.42304 | 19.7 | 2 |
| 1716.45127 | -19.4 | 2 |
| 1716.46367 | 10.4 | 2 |
| 1716.51669 | -19.1 | 2 |
| 1716.53115 | 11.5 | 2 |
| 1716.59864 | 12.5 | 2 |
| 1716.63927 | 3.2 | 2 |
| 1716.65029 | 7.5 | 2 |
| 1716.65993 | 11.2 | 2 |
| 1716.70606 | -21 | 2 |
| 1716.71226 | 6.4 | 2 |
| 1716.7405 | 17.3 | 2 |
| 1716.78732 | 10.4 | 2 |
| 1716.80316 | 16.5 | 2 |
| 1716.90232 | 4.8 | 2 |
| 1716.90232 | -20.2 | 2 |
| 1716.93262 | -8.5 | 2 |
| 1716.94846 | -2.4 | 2 |
| 1716.98496 | -13.3 | 2 |
| 1717.11167 | -14.3 | 2 |
| 1717.11304 | 11.2 | 2 |
| 1717.18466 | 13.9 | 2 |
| 1717.22047 | -22.3 | 2 |
| 1717.23769 | 9.3 | 2 |
| 1717.25559 | 16.3 | 2 |
| 1717.30862 | -13.3 | 2 |
| 1717.31412 | 13.9 | 2 |
| 1717.31688 | 14.9 | 2 |
| 1717.36922 | 10.1 | 2 |
| 1717.37335 | 11.7 | 2 |
| 1717.41191 | -23.4 | 2 |
| 1717.44565 | -10.3 | 2 |
| 1717.44978 | 16.2 | 2 |
| 1717.45598 | -6.4 | 2 |
| 1717.48077 | 3.2 | 2 |
| 1717.51314 | -9.3 | 2 |
| 1717.52416 | -5 | 2 |
| 1717.54757 | 4 | 2 |
| 1717.55377 | 6.4 | 2 |
| 1717.61506 | -19.9 | 2 |
| 1717.61712 | -19.1 | 2 |
| 1717.63778 | 13.9 | 2 |
| 1717.68805 | 8.3 | 2 |
| 1717.69081 | 9.3 | 2 |
| 1717.69838 | -12.7 | 2 |
| 1717.70182 | -11.4 | 2 |
| 1717.75072 | -17.5 | 2 |
| 1717.77757 | -7.2 | 2 |
| 1717.7879 | 21.8 | 2 |
| 1717.81063 | 5.6 | 2 |
| 1717.8368 | -9.3 | 2 |
| 1717.84919 | 20.5 | 2 |
| 1717.87536 | -19.4 | 2 |
| 1717.88224 | -16.7 | 2 |
| 1717.89877 | -10.4 | 2 |
| 1717.91943 | -2.4 | 2 |
| 1717.95042 | -15.4 | 2 |
| 1717.95524 | 11.4 | 2 |
| 1717.9821 | 21.8 | 2 |
| 1718.02823 | 14.6 | 2 |
| 1718.04407 | -4.2 | 2 |
| 1718.0544 | -0.3 | 2 |
| 1718.09847 | 16.8 | 2 |
| 1718.18524 | 0.3 | 2 |
| 1718.18524 | 25.3 | 2 |
| 1718.22518 | -9.3 | 2 |
| 1718.23827 | -4.3 | 2 |
| 1718.27683 | -14.4 | 2 |
| 1718.28096 | -12.8 | 2 |
| 1718.2844 | -11.4 | 2 |
| 1718.33054 | -18.6 | 2 |
| 1718.3333 | -17.6 | 2 |
| 1718.34638 | -12.5 | 2 |
| 1718.41938 | -9.3 | 2 |
| 1718.46276 | 7.4 | 2 |
| 1718.46827 | -15.4 | 2 |
| 1718.77471 | 2.9 | 2 |
| 1718.79261 | -15.2 | 2 |
| 1718.91588 | 7.4 | 2 |
| 1718.92414 | -14.4 | 2 |
| 1718.9269 | 11.7 | 2 |
| 1719.11352 | -16.2 | 2 |
| 1719.1748 | -17.6 | 2 |
| 1719.22989 | -21.3 | 2 |
| 1719.24573 | -15.2 | 2 |
| 1719.30771 | -16.2 | 2 |